\documentclass[prb,twocolumn,superscriptaddress,longbibliography]{revtex4-1}
\RequirePackage[mathlines]{lineno}
\usepackage{amssymb}
\usepackage{amsmath}
\usepackage{stmaryrd}
\usepackage{mathrsfs}
\usepackage{graphicx}
\usepackage[usenames,dvipsnames]{xcolor}
\definecolor{goodgreen}{rgb}{0.1,0.5,0}
\definecolor{goodred}{rgb}{0.7,0,0}
\usepackage{natbib}
\usepackage[colorlinks,urlcolor=goodgreen,citecolor=blue,linkcolor=goodred]{hyperref}

\makeatletter
\newsavebox{\@brx}
\newcommand{\llangle}[1][]{\savebox{\@brx}{\(\m@th{#1\langle}\)}%
  \mathopen{\copy\@brx\kern-0.5\wd\@brx\usebox{\@brx}}}
\newcommand{\rrangle}[1][]{\savebox{\@brx}{\(\m@th{#1\rangle}\)}%
  \mathclose{\copy\@brx\kern-0.5\wd\@brx\usebox{\@brx}}}
\makeatother

\usepackage[normalem]{ulem}

\setpagewiselinenumbers
\modulolinenumbers[5]

\begin{document}

\title{A brief review of mathematical foundation for analyzing topological characteristics of quantum electronic states and matter phases}

\author{V. Nam Do}
\email{nam.dovan@phenikaa-uni.edu.vn}    
\affiliation{Department of Basic Science, Phenikaa Institute for Advanced Studies (PIAS), Floor 25$^{th}$, A9 Building, Phenikaa University, Hanoi 10000, Vietnam}

\begin{abstract}
We briefly review the advanced mathematical language of fiber bundle structures and how they can be used to classify two-level quantum systems based on the analysis of the topological properties of their sets of state vectors. The topological classes of quantum electronic states and matter phases are characterized by topological invariants, which can be defined geometrically as the integral of differential forms on the base manifold of the fiber bundle structure. Specifically, we demonstrate that for one-dimensional systems described by the Su-Schrieffer-Heeger (SSH) model, the set of state vectors does not always have a fiber bundle structure directly on the Brillouin zone. To classify the SSH systems, we use a technique based on the concept of composite maps to decompose the set of electronic state vectors. As a result, the SSH systems are classified based on the geometrical properties of principal fiber bundles with different base manifolds.
\end{abstract}
\maketitle

\section{Introduction}\label{Sec_I}
For nearly two decades, there has been a growing use of topological invariant concepts to characterize electronic and photonic states in various media.\cite{Cayssol_2021, Price_2022, Lan_2022} This has led to the understanding of two types of semiconductors/superconductors: ordinary and topological insulators/superconductors, as well as essential features in the electronic structure of semimetals, among other examples.\cite{Bernevig_2013} Electronic states with nontrivial topological features, such as quantized states localized at the edges of quantum Hall systems, are incredibly robust and are not destroyed by perturbations, even by changing the atomic lattice of material samples.\cite{Klitzing_1980, Thouless_1982} Therefore, nontrivial topological states are expected to have the potential to create efficient qubits for quantum computers.\cite{Kitaev_2003}

Topology is a term that may be familiar to those learning mathematics but not necessarily to those studying physics, particularly in the field of condensed matter physics (CMP). Nevertheless, topology is now commonly used in CMP and has implications for modern technologies.\cite{Kitaev_2003, Gilbert_2021} Why is this? In fact, topology and related concepts are not unfamiliar to physicists working in the domains of quantum field theories and general relativity theories, as they are used to establish abstract structures of space-times and the transformation of matter fields. The birth of topology is attributed to Euler, who used graph theory to solve the famous problem of the seven bridges of K\"onigsberg in 1735.\cite{wolfram_euler} It was then systematically developed in Poincar\'e's methodology in studying a series of basic problems in 1895.\cite{britannica_poincare} Many aspects of topology were also reflected in physics through Gauss's law (1835) and Amp\`ere's law (1825). \cite{britanica_gauss_law,britanica_ampere_law} In the 20th century, since the birth of quantum mechanics, topology has been used to solve many fundamental issues in quantum mechanics, including the Dirac magnetic monopole (1931),\cite{Dirac_1931, Wu_1975a, Wu_1975b} the Aharonov-Bohm effect (1959),\cite{Aharonov_1959, Chambers_1960} and the quantum Hall effect (1980).\cite{Klitzing_1980} The mystery of these phenomena was revealed through Berry's description of adiabatic evolution of physical systems (1984). \cite{Berry_1984, Berry_1990} Through this description, it is clear that topological and geometrical structures govern the quantum world.\cite{Resta_2000, Shapere_1989}

Fundamentally, there are two problems in physics that we are concerned with. The first is to characterize the existence of a system of matter, and the second is to characterize its evolution over time. To tackle these issues, the concept of degrees of freedom is introduced as a means of parameterizing the system's states. Using phase space transforms the physical problem into a mathematical one, where each state is represented by a point in this space, and the set of states of the system is established as a domain in this phase space. However, it is essential to note that a geometric representation of physical states is inadequate since different points in phase space may correspond to the same physical state. The concept of ``equivalence" is used in mathematics to describe this situation, revealing the complex structure of the set of physical states as a set of points in phase space. Consequently, comprehending the transition of a system from one state to another necessitates consideration of the structural properties of the set of points in phase space. While calculus methods are typically used to perform specific calculations, the frequent use of calculus may cause us to overlook the natural meaning of the underlying operations. Calculus is built on the fundamental concept of continuity, which is not a natural concept but depends on the set of objects under consideration. Issues regarding the structure of sets have been noticed and studied by mathematicians from early on, leading to the formation of topology as a branch of modern mathematics. Although topology theories are usually presented at advanced levels, topology is often described as the study of the invariant properties of geometric objects under continuous deformation. While a common illustration of this is the topological similarity between a coffee cup and a doughnut or tire, this explanation is insufficient for abstract objects such as sets of physical system states. Nevertheless, the analogy of deforming one object into another may relate to the dynamics of the physical system, as the transition of a system from one state to another requires changing some parameters of the physical system. In CMP, the theory of the energy band structure of electrons in periodic lattices is a direct and significant achievement of quantum mechanics. The success of this theory allows us to distinguish between two types of materials, metals and insulators. However, this theory is built on the infinite extent of the periodic lattice, so there are difficulties in describing certain physical phenomena such as the charge polarization in the insulators. In fact, the periodicity of the atomic lattice has been used to calculate the energy band structure of electrons, which, mathematically, is a way of compactifying infinite space. When rewriting the theory of charge polarization, a quantity similar to the geometrical Berry phase appears, which is defined as the integral of a vector field along a closed path in the Brillouin zone of the crystalline lattice. This vector field is determined through the states of the electronic system and has the meaning of guiding a ``parallel motion'' within the set of electronic states.\cite{Asboth_2016} Combining all of things, one may wonder how all of these physical aspects have been resolved. Specifically, how has mathematics been applied, and what reasoning method is used here?

The purpose of this paper is to provide a brief review of the minimum basic mathematical foundation and to demonstrate the application of topological theories in condensed matter quantum physics. We employ two classical toy models, one for general two-level systems and the Su-Schrieffer-Heeger (SSH) model for one-dimensional electronic lattices, to demonstrate a procedure for topologically characterizing and classifying quantum systems. These models are commonly used in lectures and overview articles,\cite{Asboth_2016,Bernevig_2013,Cayssol_2021,Lan_2022} but they are usually presented in a mathematically loose manner. By employing precise and rigorous mathematical language to correctly state physical problems, we show that the set of state vectors does not always have the fiber bundle structure defined on a compact manifold, such as the Brillouin zone, as intuitively thought. Therefore, the analysis of the topological structure of the set of state vectors is rather tricky. We aim to present the rigorous and advanced mathematical language in a familiar way to resolve obstacles in the ``classical mindset". We highlight the interpretation of expressions such as ``investigating behaviors of a physical system" as "the investigation of features of a map" or ``a set of state vectors of a physical system" as a vector field in a space of parameters where a set of equivalent states is assigned at each point. Such a field of state vectors is determined by a continuous map from a parameter space to a special space of the fiber bundle structure. The movement in this space is guided by a quantity called the connection, which is mathematically defined as a differential form in the parameter space. The invariant characters of the set of quantum states characterizing the movement are the geometrical properties of the fiber bundle, which are usually determined by an integer number, the value of an integral of the differential form over the entire parameter space.

This article, aside from the introductory and concluding sections, is divided into two main parts. Part I, presented in Sec. \ref{Sec_II}, provides an overview of the preliminary mathematical concepts in topology that are necessary to understand and apply the methodology. Part II, consisting of Secs. \ref{Sec_III} and \ref{Sec_IV}, presents the methodology, along with detailed instructions, for analyzing two physical models, each of which is presented in a separate section.

\section{Fundamental mathematical concepts}\label{Sec_II}
The aim of this section is to present a non-exhaustive list of fundamental mathematical concepts that commonly appear in the analysis of the topological properties of physical systems. Each concept is briefly introduced, but for further details, readers are encouraged to refer to textbooks on topology such as \onlinecite{Nakahara_2003,Hassani_1999,Simon_2021}. These concepts serve as a basic vocabulary for discussions in this field.

\subsection{Topological spaces and continuous maps}\label{Sec_IIA}
{\bf Topology:} It is the primary concept used to define other fundamental mathematical concepts that describe physical objects as mathematical structures from the viewpoint of the set of elementary elements. The topology defines the neighborhood of an element in a set and the continuity in the variation of a quantity with respect to another. The formal definition of topology can be stated as follows: Given a set $X$ of objects, a topology $T_X$ of $X$ is a collection of subsets $\{U_i\}$ of $X$ such that:
\begin{enumerate}
    \item Both $X$ and the empty set belong to $T_X$.
    \item Any union of elements in $T_X$ belongs to $T_X$.
    \item A finite intersection of elements in $T_X$ belongs to $T_X$.
\end{enumerate}

{\bf Open sets:} The subsets of $X$ that belong to the topology $T_X$ are called open sets.

{\bf Topological spaces:} Sets of objects endowed with an appropriate topology are called topological spaces. A topological space is usually written as a pair $(X,T_X)$ or simply as the set $X$ if $T_X$ is a normal topology or if the context makes it clear. Each object in a topological space is called a point.

{\bf Neighborhood:} A subset $N$ of a set $X$ is called a neighborhood of a point $p\in X$ if $N$ contains at least one open set that contains the point $p$, i.e., $N\supset U_i$ such that $p\in U_i$ and $U_i\in T_X$. If $N$ is an open set, it is called an open neighborhood.

{\bf Coverings:} A collection of subsets ${U_i}$ of a set $X$ is called a covering of $X$ if $\bigcup_i U_i = X$. If all $U_i$ belong to $T_X$, then the covering is called an open covering.

{\bf Compactness:} A set is called compact if it can be covered by a finite number of open sets.

{\bf Connectedness:} A set is called connected if it cannot be partitioned into two non-empty subsets that are disjoint. In other words, the set cannot be separated into two subsets without breaking the continuity of the set.

{\bf Continuous maps:} A map $f$ between two topological spaces $X$ and $Y$ is called continuous if the preimage of an open set in $Y$ is an open set in $X$.

{\bf Homeomorphisms:} A map $f$ between two topological spaces $X$ and $Y$ is called a homeomorphism if $f$ and its inverse are both continuous. This means that there is a correspondence not only between elements in the two sets $X$ and $Y$, but also between open sets in the two topologies $T_X$ and $T_Y$. If there is a homeomorphism between two topological spaces, we say that the two topological spaces are homeomorphic to each other. A homeomorphism defines an equivalence relation on the set of all topological spaces. As a consequence, given a homeomorphism, we can classify all topological spaces into equivalence classes.

{\bf Topological invariants:} Integer numbers are often used to characterize common features of topological spaces in the same equivalence class under the homeomorphism relation. These integer values are known as topological invariants.

{\bf Homotopy and deformation:} Homotopy is a relation between two continuous functions that captures the idea of ``continuous deformation". More specifically, given two continuous functions $f, g: X \to Y$ between two topological spaces $X$ and $Y$, a homotopy between $f$ and $g$ is a continuous function $H: X \times [0,1] \to Y$ such that $H(x,0) = f(x)$ and $H(x,1) = g(x)$ for all $x$ in $X$.

\subsection{Manifolds}\label{Sec_IIB}
{\bf Manifold:} A manifold is a generalization of the concept of smooth curves and surfaces to arbitrary dimensional objects. The smoothness of a topological space allows the description of an arbitrary open neighborhood of a point by an open set in an $n$-dimensional vector space $\mathbb{R}^n$. This allows calculus on the topological space to be endowed thanks to the definition of coordinates in $\mathbb{R}^n$. Formally, we have the following definition: The topological space $\mathbb{M}$ is called a manifold if:
\begin{enumerate}
\item There exists an open covering ${U_i}$ such that $\mathbb{M} = \bigcup_i U_i$.
\item For each open set $U_i$ in the $\mathbb{M}$-covering, there exists a homeomorphism $f_i: U_i\to \mathbb{R}^n$. This map allows $U_i$ to be described by an open set in $\mathbb{R}^n$.
\item On the overlap of two open sets $U_i$ and $U_j$, the composite map $t_{ij} = f_i^{-1}\circ f_j$ is differentiable. This is the compatibility condition for the description using two different maps $f_i$ and $f_j$ in $U_i\cap U_j$.
\end{enumerate}
The pairs $(U_i,f_i)$ are called charts, and the set of all charts is called an atlas of the manifold. $U_i$ is called the neighborhood, and the map $f_i$ is called the coordinate function.

{\bf Dimension:} The dimension of the vector space $\mathbb{R}^n$ that the manifold locally resembles is called the dimension of the manifold.

{\bf Functions:} Maps $f:\mathbb{M}\to \mathbb{R}$, where $p\in \mathbb{M}\mapsto f(p) \in \mathbb{R}$, are called functions defined on the manifold. Each bijective map $f_i: \mathbb{M}\to\mathbb{R}^n$ is represented by $n$ functions $p\in U_i\mapsto f_i(p)=(x^1(p),x^2(p),\dots,x^n(p))$.

{\bf Curves:} A differentiable curve in the manifold $\mathbb{M}$ is a $C^\infty$-map of an interval of $\mathbb{R}$ to $\mathbb{M}$, i.e., $c:[a,b]\subset \mathbb{R}\to \mathbb{M}$, where $t\in [a,b]\mapsto c(t) = f_i^{-1}(x^1(c(t)),x^2(c(t)),\dots,x^n(c(t)))$.

{\bf Tangent vectors and tangent spaces:} A tangent vector is an object $V(p)$ defined at each point of the manifold $\mathbb{M}$ that acts as the derivative on functions at the point $p$, i.e., $V(p) = V^\mu(p) \partial/\partial x^\mu$, where $V[f(p)] = V^\mu(p)\partial f(p)/\partial x^\mu$. Here $p\in U_i$ and $f_i(p) = (x^1(p),x^2(p),\dots,x^n(p))$. The set of $n$ numbers $V^\mu\in \mathbb{R}$ is called the set of coordinates of the vector $V(p)$.

{\bf Tangent bundle:} The union of all tangent vectors over the manifold $\mathbb{M}$ is called the tangent bundle, $T(\mathbb{M}) = \bigcup_{p\in M}T_p(\mathbb{M})$.

{\bf Vector fields:} A vector field $V$ is a continuous map from $\mathbb{M}$ to the tangent bundle $T(\mathbb{M})$, i.e., $V:\mathbb{M}\to T(\mathbb{M})$, where $p\mapsto V(p)\in T_p(\mathbb{M})$.

{\bf One-forms and cotangent spaces:} A one-form is a linear object $\omega$ defined at each point $p\in \mathbb{M}$ that maps tangent vectors at $p$ to a number, i.e., $\langle \omega(p), V(p)\rangle \in \mathbb{R}$. The set of all possible one-forms defined at a point $p\in \mathbb{M}$ forms a vector space called the cotangent space, denoted by $T_p^\star(\mathbb{M})$. If the partial derivatives $\partial/\partial x^\mu$ are seen as the basis vectors of the tangent space, the differential $dx^\mu$ plays the role of the basis vectors of the cotangent space, i.e., $\omega(p) = \omega_\mu(p)dx^\mu$ where $\omega_\mu(p)\in \mathbb{R}$, because $\langle dx^\mu, \partial/\partial x^\nu\rangle = \delta^\mu_\nu$. Here, $\delta^\mu_\nu$ denotes the Kronecker delta symbol.

{\bf Tensor fields:} A tensor of rank $(q,r)$ is a multilinear object that takes $q$ elements of $T^\star_p(\mathbb{M})$ and $r$ elements of $T_p(\mathbb{M})$ and returns a number. The set of all tensors of rank $(q,r)$ defined at the point $p$ on the manifold $\mathbb{M}$ is denoted by $\mathcal{T}^q_{r,p}(\mathbb{M})$. The union $\bigcup_{p\in \mathbb{M}}\mathcal{T}^q_{r,p}(\mathbb{M})$ is called the tensor bundle. A tensor field of rank $(q,r)$ is a continuous map $T:\mathbb{M}\to \mathcal{T}^q_r(\mathbb{M})$.

{\bf Connection:} Intuitively, this concept is an instruction to a special motion in a manifold, i.e., the so-called parallel transport. The connection is a central quantity to describe the geometrical properties of manifolds. The definition is quite technical, so readers should consult textbooks of topology.\cite{Nakahara_2003,Hassani_1999,Simon_2021}

{\bf Curvature:} Curvature is a concept used to characterize a geometrical feature of manifolds. The general definition of this concept is quite technical, so we ask readers to consult textbooks on geometry and topology.\cite{Nakahara_2003,Hassani_1999,Simon_2021}

\subsection{Fiber bundles}\label{Sec_IIC}
{\bf Fiber bundle:} The tangent bundle and cotangent bundle are two natural geometric objects associated with a manifold that allow for the definition of vector fields on the manifold. A fiber bundle is a generalization and a natural mathematical setting to describe physical fields. Technically, it is a concept used to refer to a special kind of manifold that locally looks like the direct product of two manifolds. Conversely, we can construct a fiber bundle from some other manifolds. The formal definition is as follows: A manifold $\mathbb{E}$ is said to have a fiber bundle structure over the base manifold $\mathbb{B}$ with fiber manifold $\mathbb{F}$ if there exists a surjective map $\hat{\pi}: \mathbb{E} \rightarrow \mathbb{B}$ satisfying the following conditions:
\begin{enumerate}
\item For all points $p\in\mathbb{B}$, the preimage $\hat{\pi}^{-1}(p)$ of $p$ by the map $\hat{\pi}$ is homeomorphic to the manifold $\mathbb{F}$.
\item For each open set $U_i$ of an open covering ${U_i}$ of $\mathbb{B}$, its preimage $\hat{\pi}^{-1}(U_i)$ is simply described as a direct product $U_i\times \mathbb{F}$ by a diffeomorphism $\phi_i:U_i\times \mathbb{F}\to \hat{\pi}^{-1}(U_i)$. The maps $\phi_i$ are called the local trivializations.
\item The description of $\hat{\pi}^{-1}(U_i)$ as $U_i\times\mathbb{F}$ must be consistent. This means that the composite map $\phi_i^{-1}\circ\phi_j:(U_i\cap U_j)\times\mathbb{F}\to(U_i\cap U_j)\times\mathbb{F}$ must satisfy the condition $\phi_i^{-1}\circ\phi_j(p,f) = (p,g_{ij}(p)f)$, where $g_{ij}(p)$ are the functions determined by the map $g_{ij}:U_i\cap U_j\to \mathbb{F}$ that have the properties: $g_{ii}(p) = id_{U_i}$ and $g_{ij}(p)g_{jk}(p)g_{ki}(p) = id_{U_i}$.
\end{enumerate}
The set $\mathbb{E}$, called the total/entire space, is thus decomposed into the fibers $\mathbb{F}(p)$, i.e., $\mathbb{E}=\bigcup_{p\in\mathbb{B}}\mathbb{F}(p)$, where $\mathbb{F}(p) = \hat{\pi}^{-1}(p)$ is called the fiber attached to the base manifold $\mathbb{B}$ at the point $p$.

{\bf Principal bundles:} Principal bundles are fiber bundles in which the fiber $\mathbb{F}$ is identical to the structure group $G$. In the next section, we will work with this type of fiber bundle with the structure group $G=U(1)$ (the gauge $U(1)$ group).

{\bf Cross-sections:} Let $\hat{\pi}: \mathbb{E}\to \mathbb{B}$ be a fiber bundle. A cross-section of the fiber bundle is a smooth map $s: \mathbb{B}\to \mathbb{E}$. Clearly, $p\mapsto s(p)$ is an element of $\mathbb{F}_p=\hat{\pi}^{-1}(p)$.

{\bf Connections:} As mentioned in the previous subsection, it is rather technical to define the connection. However, because of the special structure of the fiber bundles, the idea of defining the connection is a way to separate the tangent vector of the total space into the vertical (along the fiber) and horizontal (along the base space) directions. Again, readers are asked to consult textbooks.\cite{Nakahara_2003,Hassani_1999,Simon_2021}

\section{General model for two-level systems}\label{Sec_III}
This section presents an analysis of a toy model for generic two-level systems to highlight the application of the fiber bundle structure in characterizing a set of quantum states. Mathematically, this section illustrates the analysis of a set of points defined by an appropriate map.

\subsection{Physical model}\label{Sec_IIIA}
The general model for the dynamics of an electron in a two-level system is defined by a two-dimensional Hamiltonian matrix. Based on the Hermitian property of physical observables, the Hamiltonian matrix reads
\begin{align}
H &= \begin{pmatrix}
d_0+d_z & d_x-id_y\\
d_x+id_y & d_0-d_z
\end{pmatrix} \nonumber\\
&= d_0\sigma_0+d_x\sigma_x+d_y\sigma_y+d_z\sigma_z,\label{Eq1}
\end{align}
where $\sigma_0$ is the $2\times 2$ identity matrix, and $\sigma_x,\sigma_y,\sigma_z$ are three conventional Pauli matrices. The parameters $d_x,d_y,d_z$ and $d_0$ are real and determine the properties of the system. Since the first term in Eq. (\ref{Eq1}) can be seen as the energy reference, it can be ignored in the following analysis. Therefore, the Hamiltonian matrix is determined by the remaining three terms. Defining the 3-dimensional vectors $\mathbf{d} = (d_x,d_y,d_z)$ and $\boldsymbol{\sigma} = (\sigma_x,\sigma_y,\sigma_z)$, the Hamiltonian matrix can be expressed in a compact form as the scalar product of two vectors:
\begin{align}\label{Eq2}
H = \mathbf{d}\cdot\boldsymbol{\sigma}.
\end{align}
The vector $\mathbf{d}$ represents a point in the 3-dimensional linear space $\mathbb{R}^3$. In this space, we do not consider the origin point $(0,0,0)$ as it defines a non-trivial physical system. Any other point in $\mathbb{R}^3$ defines a particular 2-level system. In other words, the existence condition of the 2-level systems is defined by the parameter vector $\mathbf{d}$.

\subsection{Determination of eigen-states}\label{Sec_IIIB}
An eigen-state of a quantum system is determined by an energy value and a set of objects known as state vectors. Mathematically, the eigen-energies and the associated eigen-state vectors are determined by the eigen-value equation of the Hamiltonian $H|\psi\rangle = E|\psi\rangle$. Specifically, the eigen-values $E$ are determined by the secular equation:
\begin{align}\label{Eq3}
&\det\begin{pmatrix}
d_z-E & d_x-id_y\\
d_x+id_y & -d_z-E
\end{pmatrix} = 0 \quad \nonumber\\
\rightarrow \quad & E^2-d_z^2-(d_x+id_y)(d_x-id_y) = 0.
\end{align}
This equation has two solutions for the unknown $E$ given by the formulae:
\begin{align}\label{Eq4}
E = \pm \sqrt{d_x^2+d_y^2+d_z^2} = \pm |\mathbf{d}| = \pm d = E_\pm(\mathbf{d}).
\end{align}
The eigen-values of the Hamiltonian $H$ depend on the Euclidean length $d$ of the vector $\mathbf{d}$. The parameter space $\mathbb{R}^3$ is therefore identified as the Euclidean space $\mathbb{R}^3$. Since the point $\mathbf{d} = (0,0,0)$ is not considered, we see that these two eigen-values are always separated from each other by a finite amount:
\begin{align}\label{Eq5}
\Delta E(\mathbf{d}) = E_+(\mathbf{d})-E_-(\mathbf{d}) = 2d > 0.
\end{align}
This confirms that the model (\ref{Eq2}) is appropriate for describing two-level systems.
To determine the eigenvectors associated with the two eigenvalues, we define the generic state vector of the two-level systems as follows:
\begin{align}\label{Eq6}
|\Psi\rangle = \begin{pmatrix}
\phi_1\\ \phi_2
\end{pmatrix},
\end{align}
where $\phi_1$ and $\phi_2$ can take on complex values. The eigenvectors of the Hamiltonian matrix are denoted by $|+,\mathbf{d}\rangle$ and $|-,\mathbf{d}\rangle$, corresponding to the eigenvalues $E_\pm(\mathbf{d})$, respectively. The equation $[H-E_-(\mathbf{d})]|-,\mathbf{d}\rangle = 0$ is specified as follows:
\begin{align}\label{Eq7}
&\begin{pmatrix}
d_z+d & d_x-id_y\\
d_x+id_y & -d_z+d
\end{pmatrix}
\begin{pmatrix}
\phi_1 \\ \phi_2
\end{pmatrix}
=\begin{pmatrix}
0 \ 0
\end{pmatrix} \nonumber\\
&\Rightarrow (d_z+d)\phi_1+(d_x-id_y)\phi_2 = 0.
\end{align}
To find $\phi_1$ and $\phi_2$, we need to consider the following cases:
(1) If $d_z = d$, then $d_x$ and $d_y$ are both zero, i.e., $d_x=d_y = 0$. In this case, $\phi_1$ and $\phi_2$ can be arbitrarily chosen, and the eigenvector is therefore ill-defined.
(2) If $d_z\neq d$, Eq. (\ref{Eq7}) allows us to find the relationship, but not a specific value, between the two components $\phi_1$ and $\phi_2$ of the eigenvector. Specifically, we can write:
\begin{align}\label{Eq8}
\left\{\begin{array}{l}\phi_1 = \gamma(-d_x+id_y)\\
\phi_2 = \gamma(d+d_z)\end{array}\right.,
\end{align}
where $\gamma$ is a non-zero undetermined complex factor. Normalizing the length of $|-,\mathbf{d}\rangle$, we calculate:
\begin{align}\label{Eq9}
|\phi_1|^2+|\phi_2|^2 = 2|\gamma|^2d(d+d_z).
\end{align}
The expression of the eigenvector $|-,\mathbf{d}\rangle$ thus reads:
\begin{align}\label{Eq10}
    |-,\mathbf{d}\rangle = \frac{e^{i\chi}}{\sqrt{2d(d+d_z)}}\begin{pmatrix}
    -d_x+id_y\\ d+d_z
    \end{pmatrix},
\end{align}
Where $e^{i\chi} = \gamma/|\gamma|$ is the argument of the complex factor $\gamma$. Here, $\chi$ is a real parameter. So, the eigen-vector $|-,\mathbf{d}\rangle$ is not uniquely defined, but is up to a phase factor. In other words, we can state that given a vector $\mathbf{d}\in\mathbb{R}^2\backslash{(0,0,0)}$, Eq. (\ref{Eq7}) does not define one state vector, but a set of state vectors that differ from each other by a $U(1)$ gauge factor $e^{i\chi}$. Physically, all state vectors in this set describe the same state of the system. Hence, they are classified into a unique equivalence class with the $U(1)$ equivalence relation, i.e., $[\![|-,\mathbf{d}\rangle_0]\!] = {g(\mathbf{d})|-,\mathbf{d}\rangle_0,|,\forall, g(\mathbf{d})\in U(1)}$, where the representative element $|-,\mathbf{d}\rangle_0$ is chosen as:
\begin{align}\label{Eq11}
|-,\mathbf{d}\rangle_0 = \frac{1}{\sqrt{2d(d+d_z)}}\begin{pmatrix}
d_x-id_y\\ d-d_z
\end{pmatrix}.
\end{align}
The discussion is totally similar to the eigen-vector $|+,\mathbf{d}\rangle$, which is given by the formula:
\begin{align}\label{Eq12}
|+,\mathbf{d}\rangle = \frac{e^{i\chi}}{\sqrt{2d(d-d_z)}}\begin{pmatrix}
d_x-id_y\\ d-d_z
\end{pmatrix}.
\end{align}

Due to the dependence of the eigen-energies and state vectors on the length of the $\mathbf{d}$ vector, we can change the parameterization of $\mathbf{d}$ from Cartesian coordinates $(d_x,d_y,d_z)$ to spherical coordinates $(d,\varphi,\theta)$, where $d>0$, $\varphi\in[0,2\pi]$, and $\theta\in[0,\pi]$. The eigen-vectors $|\pm,\mathbf{d}\rangle$ can then be rewritten in the form:
\begin{subequations}
\begin{align}
&|+,\varphi,\theta\rangle = e^{i\chi}\begin{pmatrix}
e^{-i\varphi}\cos(\theta/2)\\ \sin(\theta/2)
\end{pmatrix},\label{Eq13a}\\
&|-,\varphi,\theta\rangle = e^{i\chi}\begin{pmatrix}
e^{-i\varphi}\sin(\theta/2)\\ -\cos(\theta/2)
\end{pmatrix}.\label{Eq13b}
\end{align}
\end{subequations}
Using spherical coordinates, we find that the eigen-state vectors of the 2-level system do not depend on the length of the model parameter $\mathbf{d}$ vector, but only on the two angular coordinates. We can therefore classify the set of all possible values of the vector $\mathbf{d}$ into a unique equivalence class represented by the unit sphere $\mathbb{S}^2$ centered at the origin of the Cartesian axes in the Euclidean $\mathbb{R}^3$ space (this two-dimensional surface is embedded in the Euclidean space $\mathbb{R}^3$). Each point on the unit sphere is parameterized by only two real variables $\varphi\in[0,2\pi]$ and $\theta\in[0,\pi]$. A point on the sphere $\mathbb{S}^2$ is determined by a point $(\varphi,\theta)$ in the rectangular domain $[0,2\pi]\times[0,\pi]$. Topologically, we see that the sphere $\mathbb{S}^2$ is homeomorphic to the rectangular domain $[0,2\pi]\times[0,\pi]$.

\subsection{Investigation of topological features of quantum states}\label{Sec_IIIC}
Physically, we would like to know what happens when we drive the system. According to the model given by Eq. (\ref{Eq1}), the condition for the existence of a two-level quantum system is encoded in the parameter vector $\mathbf{d}$. Eq. (\ref{Eq4}) shows the linear dependence of the energy values of the two eigen-states of a system on the length of the $\mathbf{d}$ vector. Mathematically, the energy spectrum of a system is determined by scalar fields on the manifold of parameters. Some important features of such scalar fields can manifest through the picture of the density of states such as the van Hove singularities. However, these are local geometrical features of the energy isosurfaces, i.e., the existence of local extremal and/or saddle points of the surfaces. Driving the $\mathbf{d}$ vector on the unit sphere $\mathbb{S}^2$, the energy spectrum of the system does not change. Meanwhile, the state vectors explicitly depend on the angle coordinates of the $\mathbf{d}$ vector, see Eqs. (\ref{Eq13a}) and (\ref{Eq13b}), so they vary when driving $\mathbf{d}$ on $\mathbb{S}^2$. Accordingly, it is natural to ask: what information can the state vectors provide as the system is driven? To proceed with the discussion, we consider the set of state vectors
\begin{align}\label{Eq14}
\mathbb{E}=\left\{|-,\mathbf{d}\rangle=e^{i\chi}|-,\mathbf{d}\rangle_0,|,\forall\chi\in\mathbb{R},\forall\mathbf{d}\in\mathbb{S}^2\right\}.
\end{align}
We consider this set due to the fact that physical systems tend to stay in their lowest-energy state. The detailed discussion is presented in the following subsections.

\subsubsection{The fiber-bundle structure of the set of state vectors}\label{Sec_IIIC1}
Let us denote
\begin{align}\label{Eq15}
\mathbb{F}(\mathbf{d}) = \left\{|-,\mathbf{d}\rangle=\left.\frac{e^{i\chi}}{\sqrt{2d(d+d_z)}}\begin{pmatrix}
-d_x+id_y \\ d+d_z
\end{pmatrix},\right|,\forall\chi\in\mathbb{R}\right\}.
\end{align}
The set of state vectors $\mathbb{E}$ can thus be rewritten as:
\begin{align}\label{Eq16}
\mathbb{E} = \bigcup_{\mathbf{d}\in\mathbb{B}}\mathbb{F}(\mathbf{d}),
\end{align}
where $\mathbb{B} = \{\mathbf{d}\in\mathbb{R}^3,|, \text{norm}_2(\mathbf{d})=1\}\equiv \mathbb{S}^2$. We now show that the set $\mathbb{E}$ has a fiber bundle structure with the fiber being the manifold $\mathbb{F}=\mathbb{S}^1$ endowed by the $U(1)$ structure group.

Indeed, first of all, let us show that there exists a surjective map from $\hat{\pi}$ projecting $\mathbb{E}$ onto $\mathbb{B}$, i.e., $\hat{\pi}:\mathbb{E}\to\mathbb{B}$. To do so, we pick a generic element $|-\rangle = (\phi_1,\phi_2)^T$ in $\mathbb{E}$ and identify it with an element $|-,\mathbf{d}\rangle$ in a subset $\mathbb{F}(\mathbf{d})$. From Eq. (\ref{Eq10}), we can construct the map $\hat{\pi}$, given by the following explicit rule:
\begin{align}\label{Eq17}
\hat{\pi}:(\phi_1,\phi_2)\in \mathbb{E}\mapsto
\left\{\begin{array}{l}
d_x = -2\text{Re}\left(|\phi_2|\frac{\phi_1}{\phi_2}\right)\\
d_y = 2\text{Im}\left(|\phi_2|\frac{\phi_1}{\phi_2}\right)\\
d_z = 2|\phi_2|-1
\end{array}
\right.\in \mathbb{B}.
\end{align}
This means that the set $\mathbb{F}(\mathbf{d})$ is the preimage of $\mathbf{d}$ under the map $\hat{\pi}$, i.e., $\mathbb{F}(\mathbf{d})=\hat{\pi}^{-1}(\mathbf{d})$.

Next, we will show that the manifold $\mathbb{E}$ can be locally described as the direct product of an open set of $\mathbb{B}$ and another manifold. We first notice that the set $\mathbb{F}(\mathbf{d})$ is homeomorphic to the set $\mathbb{F}=\mathbb{S}^1$ because all pairs of complex numbers $(\phi_1,\phi_2)$ that are different from each other by a phase factor $e^{i\varphi}$ are mapped onto the same $\mathbf{d}$ point in $\mathbb{S}^2$. We also notice that $\mathbb{F}(\mathbf{d})$ is ill-defined at $\mathbf{d}=(0,0,-d)$ because of the singularity of the factor $1/\sqrt{2d(d+d_z)}$ in the expression of the state vectors. However, this singularity, as shown below, can be apparently avoided by choosing an appropriate scheme of local parameterization. Indeed, using the spherical coordinates as a specific parameterization of the $\mathbb{S}^2$ sphere seems to eliminate the singularity because of the disappearance of the factor $1/\sqrt{2d(d+d_z)}$. Actually, the condition $d_z = -d$ exactly corresponds to the value $\theta = \pi$, but $\varphi$ is still not uniquely determined. Hence, the state vector $|-,\mathbf{d}\rangle_0$ is ill-defined at the south pole of the sphere $\mathbb{S}^2$. For other state vectors, $|-,\mathbf{d}\rangle = \exp(i\chi)|-,\mathbf{d}\rangle_0$, we see that if $\chi$ is identical to $\varphi$ then:
\begin{align}\label{Eq18}
    e^{i\chi}|-,\mathbf{d}\rangle_0 =
    e^{i\varphi}\begin{pmatrix}
     e^{-i\varphi}\sin(\theta/2)\\ -\cos(\theta/2)
    \end{pmatrix} = 
    \begin{pmatrix}
    \sin(\theta/2) \\ -e^{i\varphi}\cos(\theta/2)
    \end{pmatrix}.
\end{align}
These state vectors are clearly defined at the south pole, eliminating the problem of ill-definition. However, we realize that the new state vectors become ill-defined at the north pole with $\theta = 0$ and arbitrary $\varphi$. As we can see from Eq. (\ref{Eq10}), the singularity of the map $|-\rangle$ is actually permanent and cannot be removed by choosing a particular parameterization and a gauge transformation. The gauge transformation simply moves the singularity point of the map $|-\rangle_0:\mathbb{S}^2\to \mathbb{F}(\mathbf{d})$ from one point on the domain manifold $\mathbb{S}^2$ to another point. In other words, the map $|-\rangle_0$ is only locally defined in the entire manifold $\mathbb{S}^2$.

To locally describe the manifold $\mathbb{E}$, we must use some open sets to cover the manifold $\mathbb{B}=\mathbb{S}^2$. Concretely, we use two open sets $U_N = [0,2\pi]\times[0,\pi/2+\epsilon)$ and $U_S = [0,2\pi]\times (\pi/2-\epsilon,\pi]$ to cover the north and south half spheres, respectively. The overlap of these two open sets is the ribbon covering the equator, i.e., $U_N\cap U_S = [0,2\pi]\times (\pi/2-\epsilon,\pi/2+\epsilon)$. We can then define the maps $|-\rangle_{N/S}:U_{N/S}\times U(1)\to \hat{\pi}^{-1}(U_{N/S})$ in each open set as follows:
\begin{align}
    &|-\rangle_N:(\varphi,\theta;e^{i\chi})\mapsto|-,\varphi,\theta\rangle_N = e^{i\chi}\begin{pmatrix}
    e^{-i\varphi}\sin(\theta/2) \\ -\cos(\theta/2)
    \end{pmatrix},\label{Eq19}\\
    &|-\rangle_S:(\varphi,\theta;e^{i\chi})\mapsto|-,\varphi,\theta\rangle_S = e^{i\chi}\begin{pmatrix}
    \sin(\theta/2) \\ -e^{i\varphi}\cos(\theta/2)
    \end{pmatrix}.\label{Eq20}
\end{align}
Clearly, there is no problem of singularity of these maps on each chart $U_N$ and $U_S$. These maps are diffeomorphic since their inverse maps, e.g., $|-\rangle_N^{-1}$, given by the functions:
\begin{align}\label{Eq21}
    \begin{pmatrix}
    \phi_1\\\phi_2
    \end{pmatrix}\mapsto
    \begin{pmatrix}
    -\frac{i}{2}\ln\left[\left(\frac{\phi_1}{\phi_2}\right)^2\cdot\frac{|\phi_2|^2}{1-|\phi_2|^2}\right],2\arccos{|\phi_2|};  e^{i\chi}
    \end{pmatrix},
\end{align}
is analytic. So, the pairs $(U_N,|-\rangle_N)$ and $(U_S,|-\rangle_S)$ are identified as the local trivializations.

The last point we would like to show is that the transition function $t_{NS}$ is an element of the gauge $U(1)$ group. Indeed, let $t_{NS}(\varphi,\theta)$ denote the function connecting $|-,\varphi,\theta\rangle_S$ to $|-,\varphi,\theta\rangle_N$ for $(\varphi,\theta)\in U_N\cap U_S$. From the requirement:
\begin{align}\label{Eq22}
    e^{i\chi_N}\begin{pmatrix}
    e^{-i\varphi}\sin(\theta/2)\\ -\cos(\theta/2)
    \end{pmatrix} = t_{NS}(\varphi,\theta)e^{i\chi_S}\begin{pmatrix}
    \sin(\theta/2) \\ -e^{i\varphi}\cos(\theta/2)
    \end{pmatrix},
\end{align}
it is easy to deduce the identification:
\begin{equation}\label{Eq23}
    t_{NS}(\varphi,\theta) = e^{-i\varphi+\Delta\chi_{NS}},
\end{equation}
where $\Delta\chi_{NS} = \chi_N-\chi_S$. Clearly, $t_{NS}(\varphi,\theta)$ belongs to $U(1)$ as expected. 

To sum up, we have shown that the set $\mathbb{E}$ of state vectors can be represented as a sphere $\mathbb{S}^3$ embedded in the space $\mathbb{C}^2$. This set has a fiber bundle structure with the base manifold $\mathbb{B}=\mathbb{S}^2$ and the fiber $\mathbb{F}=\mathbb{S}^1$ with the structure group $U(1)$. The projection map $\hat{\pi}:\mathbb{S}^3\to\mathbb{S}^2$ is given by Eq. (\ref{Eq17}). The local trivializations of the bundle are given by the pairs $(U_N,|-\rangle_N)$ and $(U_S,|-\rangle_S)$, which are identified on the overlap $U_N\cap U_S$ by the transition function $t_{NS}(\varphi,\theta)=\exp(i\varphi)$, which belongs to the gauge group $U(1)$.

\subsubsection{The state transition as the parallel motion in the principal bundle}\label{Sec_IIIC2}
Physical processes always involve the transition of a system from one state to another. The transition induced by varying the parameter vector $\mathbf{d}$ is mathematically translated into movement from one point to another in the fiber bundle $(\hat{\pi}:\mathbb{E} \to \mathbb{B},\mathbb{F},U(1))$. While moving in a flat space, a velocity field is enough to guide the motion, but in a curved manifold, another quantity, named the connection, is needed to keep the motion according to the curvature of the manifold, i.e., the parallel motion. The parallel motion in the fiber bundle structure can be described as the horizontal lift of the parallel motion in the base space. This description highlights the holonomy, a typical geometrical phenomenon in nontrivial fiber bundles.\cite{Nakahara_2003,Micheal_2020,Simon_2021} To determine the connection $\mathcal{A}_-$ in a fiber bundle, we need a vector field defined as a local cross-section of the bundle and then track its variation along some smooth curves. A concrete vector field $|-,\mathbf{d}\rangle_{N/S}$, locally defined on the covering $U_{N/S}$, i.e., $|-\rangle_{N/S}:(\varphi,\theta)\in U_{N/S}\mapsto e^{i\chi(\varphi,\theta)}|-,\varphi,\theta\rangle_{0,N/S}$, is determined when a smooth function $\chi(\varphi,\theta)$ on $U_{N/S}$ is given. Due to the scalar product of the state vectors defined in the Hilbert space, the so-called Berry connection is determined as follows:\cite{Micheal_2020,Bernevig_2013}
\begin{align}\label{Eq24}
    \mathcal{A}_{-}(\mathbf{d}) = i\langle -,\mathbf{d}|\hat{d}|-,\mathbf{d}\rangle,
\end{align}
where $\hat{d}$ denotes the exterior derivative \cite{Nakahara_2003}. Using the parameterization of spherical coordinates $(\varphi,\theta)$, we obtain:
\begin{align}\label{Eq25}
    \mathcal{A}_{-}(\varphi,\theta) &=  i\langle -,\varphi,\theta|\partial_\varphi|-,\varphi,\theta\rangle d\varphi \nonumber\\
    &+i\langle -,\varphi,\theta|\partial_\theta|-,\varphi,\theta\rangle d\theta.
\end{align}
On each chart $U_{N,S}$ partially covering $\mathbb{S}^2$ we obtain:
\begin{align}
    &\partial_\varphi|-,\varphi,\theta\rangle_N = e^{i\chi}\begin{pmatrix}
    -ie^{-i\varphi}\sin(\theta/2)\\0
    \end{pmatrix},\label{Eq26}\\
    &\partial_\theta|-,\varphi,\theta\rangle_N = e^{i\chi}\frac{1}{2}\begin{pmatrix}
    -ie^{-i\varphi}\cos(\theta/2)\\ \sin(\theta/2)
    \end{pmatrix}.\label{Eq27}
\end{align}
Then, it yields the expression of the connection:
\begin{align}\label{Eq28}
    \mathcal{A}_{-}(\varphi,\theta)_N = \sin^2(\theta/2)d\varphi = \frac{1}{2}(1-\cos\theta)d\varphi.
\end{align}
Similarly, we have:
\begin{align}\label{Eq29}
    \mathcal{A}_{-}(\varphi,\theta)_S = -\cos^2(\theta/2)d\varphi = -\frac{1}{2}(1+\cos\theta)d\varphi.
\end{align}
We see that the connection is not uniquely defined, but it is associated with each vector field under consideration. Since the vector fields are not globally defined, the connection in each covering of the base manifold also takes on its own form. However, in the overlap region $U_N\cap U_S$, the local connections should be related to each other. It is easy to verify that $\mathcal{A}_-(\varphi,\theta)N = \mathcal{A}_-(\varphi,\theta)_S+d\varphi$. This relation does not depend on $d\chi$ but only on $d\varphi$, as it is actually the consequence of the gauge $U(1)$ transformation.\cite{Nakahara_2003} This relation therefore ensures the existence of a globally defined connection one-form on the whole fiber bundle.

\subsubsection{Topological characters of the set of state vectors}\label{Sec_IIIC3}
The topological features of the set of state vectors $\mathbb{E}$ of two-level systems are mathematically translated into the global geometrical features of the fiber bundle $(\hat{\pi}:\mathbb{E}\to\mathbb{B},\mathbb{F},U(1))$. The analysis presented in the previous subsection diagnoses such features: the problem of singularity of the vector fields on the base manifold $\mathbb{B}\equiv\mathbb{S}^2$ of a fiber-bundle structure. This feature is quantitatively characterized by an index that is defined through an integral over the base manifold. Since the base manifold is a two-dimensional surface, we need a differential 2-form that is determined from the 1-form $\mathcal{A}_{-}$ as follows:\cite{Nakahara_2003}
\begin{align}\label{Eq30}
    \mathcal{F}_-(\varphi,\theta) &= d\mathcal{A}_-(\varphi,\theta)_{N}=d\mathcal{A}_-(\varphi,\theta)_S \nonumber\\
    &= \partial_\theta\sin^2(\theta/2)d\theta\wedge d\varphi = \frac{1}{2}\sin\theta d\theta\wedge d\varphi,
\end{align}
where ``$\wedge$" denotes the wedge product of two basis one-forms $d\theta$ and $d\varphi$ such that $d\theta\wedge d\varphi = -d\varphi\wedge d\theta$. The 2-form $\mathcal{F}_-(\varphi,\theta)$ is an antisymmetric 2-rank tensor and does not depend on $\chi$ due to the identity $d^2\chi = 0$. Geometrically, this tensor determines the local curvature of the total manifold $\mathbb{E}\equiv\mathbb{S}^3$. Therefore, the total curvature of the fiber-bundle $\mathbb{E}$ is obtained by integrating the local curvature over the entire base manifold:
\begin{align}\label{Eq31}
    \mathcal{C}_- &= \frac{1}{2\pi}\int_{\mathbb{S}^2}\mathcal{F}^- = \frac{1}{2}\int_0^{2\pi}\int_0^\pi\frac{1}{2}\sin\theta d\theta\wedge d\varphi\nonumber\\
    &= \frac{1}{4\pi}\int_0^\pi\sin\theta d\theta\int_0^{2\pi}d\varphi = -1.
\end{align}
This non-zero integer value of the integral characterizes the fact that the vector field $|-,\mathbf{d}\rangle$ is not globally defined on the manifold $\mathbb{S}^2$, but rather locally in each covering. The difference in the connection $\mathcal{A}-$ on each covering directly results in the non-zero value of the integral. Geometrically, we can understand this value as follows: by moving around the whole manifold $\mathbb{S}^2$ of the parameter vector $\mathbf{d}$ along the positive direction, the maps $|-\rangle_{N/S}$ allow us to move around the whole sphere $\mathbb{E}\equiv\mathbb{S}^3$ in one round, but via the negative direction. Thus, $\mathcal{C}_-$ (usually called the Chern number) plays the role of the winding number characterizing the topological features of the set of state vectors $\mathbb{E}$ of all two-level systems described by the model (\ref{Eq1}).

\section{Su-Shrieffer-Heeger model for one-dimensional lattices}\label{Sec_IV}
In this section, we illustrate the analysis of deforming the band structure of an atomic chain. We also demonstrate that the application of the fiber bundle structure can be flexible and somewhat complex. Specifically, we will show that the set of electronic state vectors of the atomic lattice does not always possess the fiber bundle structure directly on the Brillouin zone.

\subsection{Physical model}\label{Sec_IVA}
\begin{figure}\centering
\includegraphics[clip=true,trim=1cm 1cm 1cm 2cm,width=\columnwidth]{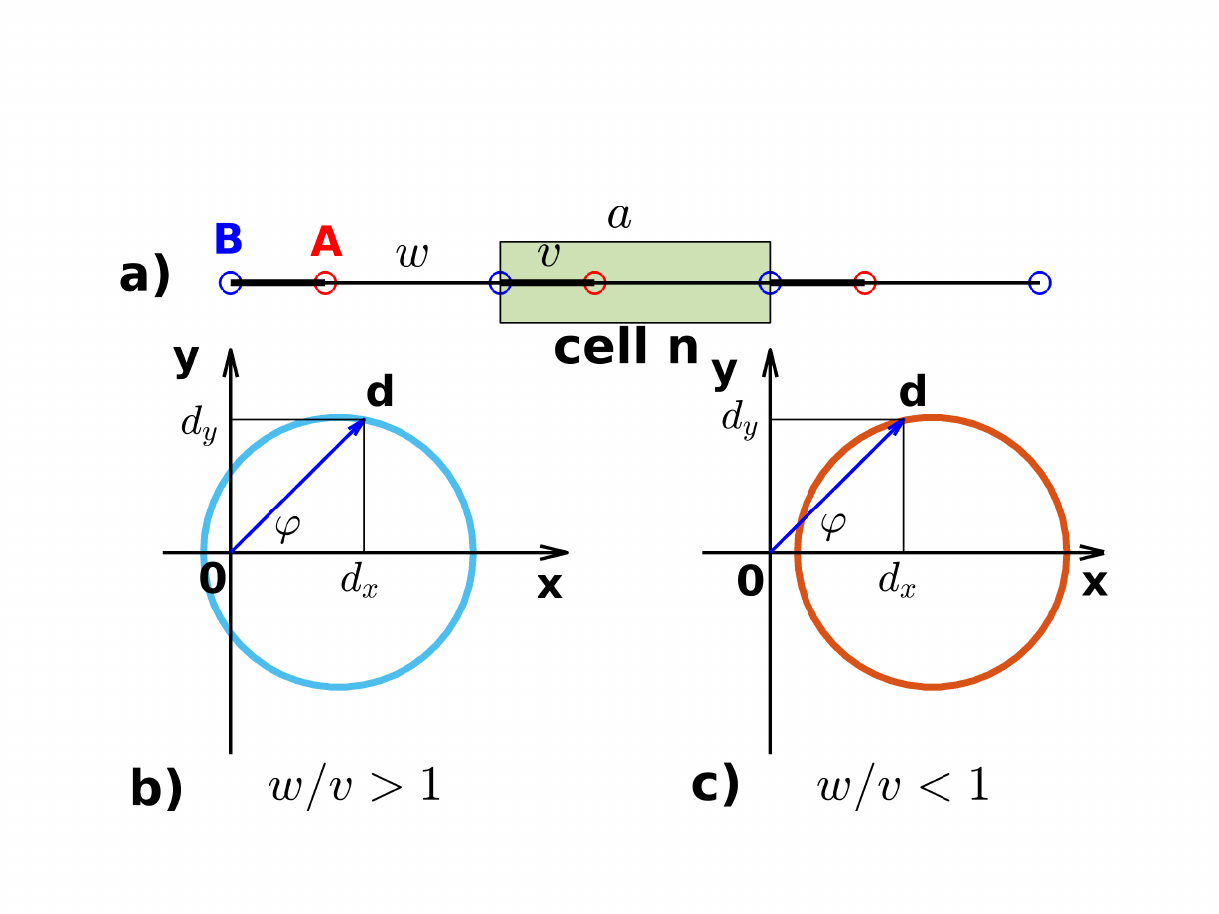}
\caption{\label{Figure_1} (a) The figure shows a one-dimensional lattice with two sub-lattices labeled $A$ and $B$, with lattice constant $a$. Each lattice node has one electronic orbital, and the kinetic energy of the electron is characterized by two hopping parameters, $v$ and $w$, denoted in the figure. The analysis presented in the text is consistent with the chosen unit cell. (b) and (c) The images of the Brillouin zone $BZ=[-\pi,\pi]$ through the map given by Eq. (\ref{Eq36}) in the cases where $w/v > 1$ and $w/v < 1$, respectively. In the former case, the polar angle $\varphi$ can take on all values in the range $[-\pi,\pi]$, while in the latter case it takes on values in the range $[-\varphi_0,\varphi_0]$, where $\varphi_0 = \arcsin(w/v)$.}
\end{figure}

Consider a one-dimensional lattice defined by two parameters $v$ and $w$, which represent the hopping energies between the two nearest lattice nodes. The lattice is periodic, with a unit cell containing two lattice nodes labeled as $A$ and $B$. Let $a$ be the length of the unit cell. Assume that each lattice node has only one electron orbital, denoted as $|\alpha,n\rangle$ for node $\alpha$ ($\alpha = A, B$) in cell $n$, as shown in Fig. \ref{Figure_1}. The Hamiltonian of an electron in the lattice is given by the following tight-binding model:
\begin{align}
    H &= \sum_n|A,n\rangle \left(v\langle B,n|+w\langle B,n+1|\right)\nonumber\\
    &+\sum_n|B,n\rangle \left(v\langle A,n|+w\langle A,n-1|\right).
\end{align}
Using the Bloch theorem we can construct the so-called Bloch state vectors $|\alpha,k\rangle$ for each value of $k$ in the Brillouin zone $BZ = \{k \in\mathbb{R}\,|\, -\pi/a \leq k\leq \pi/a\} =[-\pi/a,\pi/a]$:
\begin{align}\label{Eq32}
    &|\alpha,k\rangle = \frac{1}{\sqrt{N}}\sum_{n}e^{ikan}|\alpha,n\rangle,
\end{align}
Conversely,
\begin{align}\label{Eq33}
    |\alpha,n\rangle = \frac{1}{\sqrt{N}}\sum_{k}^{BZ}e^{-ikan}|\alpha,k\rangle,
\end{align}
Substitute Eq. (\ref{Eq33}) into the tight-binding Hamiltonian we get:
\begin{align}\label{Eq34}
    H = \sum_{k}^{BZ}\sum_{\alpha,\beta}^{\{A,B\}}|\alpha,k\rangle H_{\alpha\beta}(k)\langle \beta,k|,
\end{align}
where $H_{\alpha\beta}(k)$ are the elements of the so-called Bloch-Hamiltonian matrix that takes the following form:
\begin{align}\label{Eq35}
    H(k) = \mathbf{d}(k)\cdot\boldsymbol{\sigma}.
\end{align}
Here the dependence on $k$ is through a 2D vector $\mathbf{d}$:
\begin{align}\label{Eq36}
    \mathbf{d}(k) = \left\{\begin{array}{l}
    d_x(k) = v+w\cos(ka)\\
    d_y(k) = w\sin(ka)\\
    \end{array}\right.
\end{align}
The Bloch-Hamiltonian is defined by three parameters $v,w$ and $ka$. While the latter is real and given in the Brillouin zone $BZ = [-\pi,\pi]\simeq \mathbb{S}^1$, the two former parameters can take complex-values. For simplicity, we assume here that both $v,w$ are positive real parameters. We need to distinguish the role of these parameters: $v$ and $w$ define the physical system, and $k$ defines the state space of the physical system; which is why we denote explicitly the dependence of the Hamiltonian and the vector $\mathbf{d}$ on $k$.

\subsection{Determination of eigen-states}\label{Sec_IVB}
The eigenvalues of the Hamiltonian can be obtained through the diagonalization procedure. We can express the two eigenvalues $E_\pm(k)$ as follows:
\begin{align}\label{Eq37}
    E_\pm(k) = \pm\sqrt{d_x^2(k)+d_y^2(k)} = \pm\|\mathbf{d}(k)\| = \pm d(k),
\end{align}
where $d(k)=\sqrt{v^2+w^2+2vw\cos(kb)}$. Knowing the eigen-values, the corresponding eigen-vectors are determined by the equation $(H-E)|\psi\rangle = 0$, which is specified as:
\begin{align}\label{Eq38}
    \begin{pmatrix}
    \pm d(k) & d_x(k)-id_y(k)\\
    d_x(k)+id_y(k) & \pm d(k)
    \end{pmatrix}\begin{pmatrix}
    \phi_1(k)\\ \phi_2(k)
    \end{pmatrix} = \begin{pmatrix}
    0\\0
    \end{pmatrix}.
\end{align}
This leads to the equation:
\begin{equation}\label{Eq39}
    \pm d(k)\phi_1(k)+[d_x(k)-id_y(k)]\phi_2(k) = 0.
\end{equation}
To solve this equation, we need to consider two cases: (1) If $d(k) = 0$, it leads to $d_x(k) = d_y(k) = 0$. So, $\phi_1(k)$ and $\phi_2(k)$ can be arbitrarily chosen, and the state vectors are therefore ill-defined. Notice that in Sec. \ref{Sec_II}, we did not consider the case of $d = 0$ because the $\mathbf{d}$ vector determines the existence conditions of a particular physical system directly. Here, the two parameters $v$ and $w$ play this role, while the parameter $k$ determines the states of a particular 1D lattice. Therefore, $d(k) = 0$ can occur for a given physical system (i.e., with a given set of $v$ and $w$) for a value of $k$ in the Brillouin zone. With $d_x(k)$ and $d_y(k)$ given by Eq. (\ref{Eq36}), the case $d(k) = 0$ implies that:
\begin{align}\label{Eq40}
    \left\{\begin{array}{l} v+w\cos(ka) = 0\\ w\sin(ka) = 0\end{array} \right.\to \left\{\begin{array}{l} v=w\\ 
    ka = \pm\pi\end{array}\right.
\end{align}
It shows that $d(k) = 0$ occurs only for the configuration with $v=w$ at $ka=\pm \pi$, i.e., at the edges of the Brillouin zone.
(2) If $d(k)\neq 0$, we obtain the following expressions for the state vectors:
\begin{align}\label{Eq41}
    |\pm,k\rangle = \frac{e^{i\chi}}{\sqrt{2}}\begin{pmatrix}
    \pm \frac{d_x(k)-id_y(k)}{d(k)}\\ 1
    \end{pmatrix},
\end{align}
where $\chi$ is an arbitrary real parameter defining the gauge $U(1)$ factor. Since $\mathbf{d}(k) \in \mathbb{S}^1$ (a circle of radius $w$ embedded in the plane $\mathbb{R}^2$ at the point $(v,0)$ as the center), it can be parameterized by only one real parameter. There are many ways to parameterize the circle $\mathbb{S}^1$, but, as we will see, the use of polar angle coordinates $(d,\varphi)$ is more useful. Accordingly, $d_x = d\cos\varphi,d_y=d\sin\varphi$. From Eq. (\ref{Eq36}) we deduce the equation expressing the constraint of $d$ and $\varphi$ (see Fig. \ref{Figure_1}):
\begin{align}\label{Eq42}
    &(d\cos\varphi-v)^2+w^2\sin^2\varphi = w^2\nonumber\\
    \to & d^2-2dv\cos\varphi+(v^2-w^2) = 0.
\end{align}
Using this parameterization, we can rewrite the state vector $|\pm,\mathbf{d}(k)\rangle$ as:
\begin{align}\label{Eq43}
    |\pm,k\rangle = \frac{e^{i\chi}}{\sqrt{2}}\begin{pmatrix}
    \pm e^{-i\varphi(k)}\\ 1
    \end{pmatrix}.
\end{align}
This result shows that the state vectors do not depend on the radial coordinate $d$, but only on the polar angle coordinate $\varphi$, which is determined as a function of $k\in BZ$, see Eq. (\ref{Eq47}).

Normally, with the solution for the eigen-states, observable quantities that determine physical properties of a system can be calculated. However, as mentioned in the previous section, we are interested in what happens when we drive the physical system from one state to another, i.e., vary the state parameter $k$ in the Brillouin zone. The answer to this question is presented in the next subsection.

\subsection{Investigation of topological features of quantum states}\label{Sec_IVC}
\subsubsection{The fiber-bundle structure of the set of state vectors}\label{Sec_IVC1}
In Subsection \ref{Sec_IIB}, we consider the set of state vectors for all possible two-level systems. In contrast, in this subsection, we are interested in the features of the set of state vectors for a generic 1D SSH system. We will then compare the features of such sets to classify 1D systems described by the SSH model into different classes.

Let's denote 
\begin{align}\label{Eq44}
    \mathbb{F}(k)=\left\{|-,k\rangle =\frac{e^{i\chi}}{\sqrt{2}}\begin{pmatrix}
    -\frac{d_x(k)-id_y(k)}{d(k)}\\ 1
    \end{pmatrix}\,|\, \forall\,\chi\in\mathbb{R}\right\},
\end{align}
The set of state vectors describes the same eigen-state with the energy $E(k)=-d(k)$, as shown in Eq. (\ref{Eq37}). It is important to note that the set $\mathbb{F}(k)$ is not defined for $k$ such that $d(k)=0$, as stated in Eq. (\ref{Eq41}). This condition is satisfied only for configurations with $v=w$ at $k=\pm\pi/a$, as shown in Eq. (\ref{Eq40}).

For configurations where $v\neq w$, the sets $\mathbb{F}(k)$ are well-defined for all $k\in BZ$. Therefore, we can consider the set
\begin{align}\label{Eq45}
    \mathbb{E}=\bigcup_{k\in \mathbb{B}}\mathbb{F}(k),
\end{align}
where $\mathbb{B}=BZ\simeq \mathbb{S}^1\subset\mathbb{R}^2$. Since $\mathbb{F}(k)$ is the set of all physically equivalent state vectors, $\mathbb{E}$ mathematically represents the set of equivalent classes. To decompose $\mathbb{E}$ according to $\mathbb{B}$, we need to determine a surjective map $\hat{\pi}:\mathbb{E}\to \mathbb{B}$. To do so, we take an element $|-\rangle = (\phi_1,\phi_2)^T$ in $\mathbb{E}$ and identify it with one element in a subset $\mathbb{F}(k)$:
\begin{align}\label{Eq46}
    \left\{\begin{array}{l}
    \phi_1 = -\frac{e^{i\chi}}{\sqrt{2}}\frac{d_x(k)-id_y(k)}{d(k)}\\ \phi_2 = \frac{e^{i\chi}}{\sqrt{2}}
    \end{array}\right.
\end{align}
It results in the equation for $kb$:
\begin{align}\label{Eq47}
    \sin(ka+\varphi) = -\frac{v}{w}\sin(\varphi),
\end{align}
where $\varphi = \arg(\phi_1/\phi_2)\in [-\pi,\pi]$. It is clear that this equation does not always have a solution for $kb$ because of the factor $v/w$. Specifically, there are two cases to consider:
\begin{enumerate}
    \item $v/w < 1$: Eq. (\ref{Eq47}) always has a solution for $ka$ for all $\varphi\in [-\pi,\pi]$. It means that when taking an arbitrary element of $\mathbb{E}$, we can always identify it as belonging to a subset $\mathbb{F}(k)$. This identification procedure defines the map $\hat{\pi}:\mathbb{E}\to\mathbb{B}$, which is given by the rule: $ka = -\varphi-\arcsin(v/w\sin(\varphi))$.
    \item $v/w > 1$: Eq. (\ref{Eq47}) only has a solution for elements $(\phi_1,\phi_2)$ such that $|\sin(\varphi)| \leq w/v < 1$. In other words, the map $\hat{\pi}:\mathbb{E}\to \mathbb{B}$ is not entirely, or globally, defined in $\mathbb{B}$.
\end{enumerate}
Therefore, based on this rough analysis, 1D SSH systems can be classified into three categories according to the values of the ratio $v/w$, i.e., (1) $v/w = 1$, (2) $v/w < 1$, and (3) $v/w > 1$. The first case is special because the two energy bands $E_-(k)$ and $E_+(k)$ touch each other at the two edge points of the Brillouin zone, and the energy value $E=0$ is degenerate. Physically, the SSH configuration with $v=w$ behaves as a semi-metallic system.

To quantify the classification of the second and third cases, we move away from considering the set $\mathbb{E}$ given by Eq. (\ref{Eq45}) and instead consider:
\begin{align}\label{Eq48}
    \mathbb{E} = \bigcup_{\varphi\in \mathbb{B}}\mathbb{F}(\varphi),
\end{align}
where $\mathbb{B}$ is any open set in the interval $[-\pi,\pi]$ and
\begin{align}\label{Eq49}
    \mathbb{F}(\varphi) = \left\{
    |-,\varphi\rangle = \left.\frac{e^{i\chi}}{\sqrt{2}}\begin{pmatrix}
    -e^{-i\varphi} \\ 1
    \end{pmatrix}\,\right|\,\forall\chi\in\mathbb{R}
    \right\}.
\end{align}

To decompose the set $\mathbb{E}$ according to $\mathbb{B}$ we determine a map $\hat{\pi}:\mathbb{E}\to \mathbb{B}$. It is easy to find:
\begin{align}\label{Eq50}
    \hat{\pi}:(\phi_1,\phi_2)\mapsto \varphi = i\ln\left(-\frac{\phi_1}{\phi_2}\right).
\end{align}
This map is surjective since it maps all pairs of complex variables $(\phi_1,\phi_2)$ and $(\phi_1^\prime,\phi_2^\prime) = e^{i\chi}(\phi_1,\phi_2)$ for all $\chi\in\mathbb{R}$ to the same value of $\varphi$. It means that, given a value of $\varphi$, its preimage $\hat{\pi}^{-1}(\varphi) = \mathbb{F}(\varphi) = \mathbb{S}^1$. The fiber bundle structure of $\mathbb{E}$ is confirmed by a trivialization map $\phi:\mathbb{B}\times \mathbb{S}^1\to\hat{\pi}^{-1}(\mathbb{B})$. This map is easily defined by the assignment:
\begin{align}\label{Eq51}
\phi:(\varphi,e^{i\chi})\mapsto|-,\varphi\rangle = \frac{e^{i\chi}}{\sqrt{2}}\begin{pmatrix}
-e^{-i\varphi} \\ 1
\end{pmatrix}.
\end{align}
It is a diffeomorphism because its inverse $\phi^{-1}:\hat{\pi}^{-1}(\mathbb{B})\to\mathbb{B}\times\mathbb{S}^1$ determined by the decomposition $\phi^{-1}(|-,\varphi\rangle)=(\varphi,e^{i\chi})$ also differentiable. Note that, in the problem under consideration, we do not need a local description of the set $\mathbb{E}$ but a global one. We thus conclude that the set $\mathbb{E}$ is a fiber bundle over the base manifold $\mathbb{B}$ with the fiber $\mathbb{F}=\mathbb{S}^1\simeq U(1)$ according to the decomposition map $\hat{\pi}$.

\begin{figure}\centering
\includegraphics[clip=true,trim=0cm 0cm 0cm 0cm,width=\columnwidth]{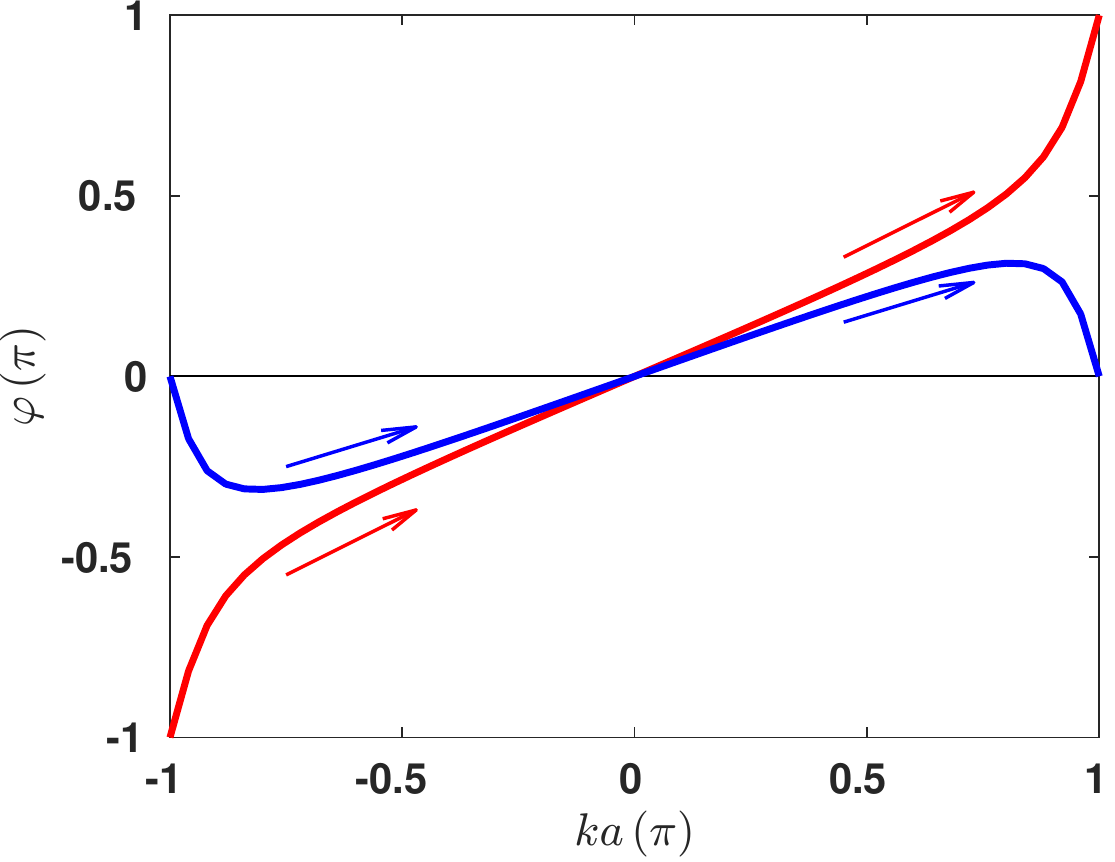}
\caption{\label{Figure_2} The variation of $\varphi$ with respect to $ka$ is shown. The blue and red curves correspond to the cases $v/w = 0.8 < 1$ and $v/w = 1.25 > 1$, respectively.}
\end{figure}

The issue now is to determine the set $\mathbb{B}$ as the image of the Brillouin zone $BZ$ through some map $\varphi:BZ\to \mathbb{B}=\varphi(BZ)$. From Eq. (\ref{Eq36}) we have:
\begin{align}\label{Eq52}
\varphi = \arg\left[v+w\cos(ka)+iw\sin(ka)\right].
\end{align}
We see that depending on the value of $v/w$ the range of $\varphi$ is determined as follows (see Fig. \ref{Figure_2}):
\begin{enumerate}
    \item $v/w > 1$: $\varphi\in [-\varphi_0,\varphi_0]\equiv\varphi(BZ)\subset [-\pi,\pi]$, where $\varphi_0 = \arcsin(w/v)$.
    \item $v/w < 1$: $\varphi\in [-\pi,\pi]\simeq \mathbb{S}^1$.
\end{enumerate}

To summarize, we distinguish two different fiber bundle structures for the set of state vectors $\mathbb{E}$. One has the base manifold $\mathbb{B} = [-\pi,\pi]$ (for $v/w < 1$) homeomorphic to the circle $\mathbb{S}^1$, and the other has the base manifold $\mathbb{B} = [-\varphi_0,\varphi_0]$ (for $v/w > 1$) simply a part of the interval $[-\pi,\pi]$.

\subsubsection{The connection of the fiber bundle}\label{Sec_IVC2}
After the determination of the fiber bundle structure of the set of state vectors $\mathbb{E}$ we can proceed to analyze its geometrical features by performing the ``exploration travelling'' in $\mathbb{E}$. Because of the principal fiber bundle structure, the parallel motion in $\mathbb{E}$ is realized by a horizontal lift of a parallel motion in the base manifold $\mathbb{B}$ up to $\mathbb{E}$. So, we need to perform a motion in $\mathbb{B}$ first. We consider separately two cases:
\begin{enumerate}
    \item $v/w > 1$: When varying $k$ from $-\pi/a$ to $\pi/a$, $\varphi$ continuously moves from $0$ to $-\varphi_0$ then back to $0$. After that it moves from $0$ to $\varphi_0$ and then to $0$. The entire path that $\varphi$ draws is $\varphi: 0\to-\varphi_0\to 0\to\varphi_0\to 0$. This path can be seen as consisting of two loops of opposite direction, one is $\varphi: 0\to\varphi_0\to 0$ with the positive direction, and the other is $\varphi:0\to-\varphi_0\to 0$ with the negative direction.
    \item $v/w < 1$: When varying $k$ from $-\pi/b$ to $\pi/a$, $\varphi$ continuously moves from $-\pi$ to $\pi$, drawing a loop homotopic to the circle $\mathbb{S}^1$.
\end{enumerate}
In Fig. \ref{Figure_2} we graphically illustrate the range of $\varphi$ and movement of $\varphi$ with respect to $ka\in[-\pi,\pi]$ according to these two cases.

After choosing a vector field as a smooth map $|-\rangle:\mathbb{B}\to \mathbb{E}$, given by $|-\rangle:\varphi\mapsto|-,\varphi\rangle$, we can determine a connection in order to horizontally lift the motion in $\mathbb{B}$ up to $\mathbb{E}$. The so-called Berry connection is then given by $\mathcal{A}- =i\langle -,\varphi|\hat{d}|-,\varphi\rangle$, where $\hat{d}$ denotes the exterior derivative. Using Eq. (\ref{Eq36}), we can write:
\begin{align}\label{Eq53}
\hat{d}|-,\varphi\rangle = \frac{e^{i\chi}}{\sqrt{2}}\begin{pmatrix}
ie^{-i\varphi}\\ 0
\end{pmatrix}d\varphi,
\end{align}
which allows us to obtain the expression for the Berry connection:
\begin{equation}\label{Eq54}
    \mathcal{A}_-(\varphi) = \frac{1}{2}d\varphi.
\end{equation}

\subsubsection{Topological characters}\label{Sec_IVC3}
Some topological features of the set of state vectors $\mathbb{E}$ are characterized by the global geometrical features of the fiber-bundle $(\hat{\pi}:\mathbb{E}\to\mathbb{B},\mathbb{F},U(1))$. Since the base manifold $\mathbb{B}$ is one-dimensional, all the higher-rank differential forms deduced from the one-form $\mathcal{A}_-$ vanish, e.g., $d\mathcal{A}_- = 0$. A topological quantity can be defined as an integral of $\mathcal{A}_-$ over the whole base manifold $\mathbb{B}$ to characterize the geometrical properties of the manifold $\mathbb{E}$, as follows:
\begin{align}\label{Eq55}
    \gamma = \int_{\mathbb{B}}\mathcal{A}_- = \left\{\begin{array}{ll}
        0 & \frac{v}{w} > 1,\\
        \pi & \frac{v}{w} < 1.
    \end{array}\right.
\end{align}
This integral is physically named the Zak phase, which determines the phase accumulated by the state vector when it completes its motion in the fiber bundle $\mathbb{E}$ according to the motion of $k$ in the entire Brillouin zone. The values of the Zak phase are the topological characteristic of the three categories of 1D lattices. If we consider the energy band structure, we realize only two phases: the semiconducting phase with a finite band gap and the semimetallic phase with a zero-band gap. However, by looking deeply into the set of state vectors, we realize that the semiconducting phase should be classified further into two different categories, characterized by two different values of the Zak phase.

\subsubsection{A view from symmetry as global constraints}
Assuming that the two parameters $v$ and $w$ are real, the Hamiltonian of the SSH atomic chain does not change under time-reversal operations. This is known as time-reversal symmetry, and it implies that the eigenvalues of the Bloch-Hamiltonian matrix $H(k)$ are even functions of $k$, as shown in Eq. (\ref{Eq37}). The time-reversal symmetry plays this role specifically for the SSH model. If we assume further that the hopping parameters $v$ and $w$ are equal, then the Hamiltonian also possesses an inversion symmetry. Together with the chiral symmetry of the Bloch-Hamiltonian matrix due to the nearest-neighbor approximation for the hopping, we find a zero-energy state with a doubly degenerate band structure. This means that the SSH atomic chain is a metal with the conduction and valence bands touch each other at $k=0$. The inversion symmetry is rigorously enforced for all possible configurations of the chain as long as $v=w$. The touching point of two energy bands is a consequence of this symmetry. It is therefore said to be protected by inversion symmetry. Breaking this inversion symmetry simply requires deforming the atomic chain such that $v\neq w$. In this case, the touching point of two energy bands is broken as expected, an energy gap appears, and the system becomes an insulator. With only two parameters, $v$ and $w$, there are two distinct insulating phases by setting $v/w < 1$ and $v/w > 1$. The question is whether these two insulating phases are equivalent or have any distinguishing features. The answer, as shown in the previous sections, is that there is a difference, and it is expressed in the structure of the set of state vectors of the system. This difference is characterized by a topological invariant quantity named the Zak phase.

\section{Conclusion}\label{Sec_V}
Topology is a fundamental concept that not only establishes the foundation for mathematical theories but also enables physicists to analyze the structure of the physical world, including space-time. However, the abstract nature and high degree of generalization of topology often pose challenges when applying topological concepts and related methods to analyze physical problems, particularly in the domain of condensed matter physics where the practical application is of high interest. Therefore, it is crucial to use mathematical concepts and rigorous mathematical language accurately to state physical problems when solving problems. We have carefully selected and presented a concise list of essential mathematical concepts as the vocabulary to discuss this topic. We have demonstrated that topological methods can be used to study problems induced by driving a quantum system through a systematic procedure. To illustrate this idea, we have analyzed two classical toy models.

The first model is general for two-level quantum systems, and the second model is for one-dimensional crystalline lattices consisting of two sub-lattices. The analysis procedure starts with determining the system's states as the objects under study. While the energies associated with quantum states are realized as scalar fields defined on a manifold of the model parameters, the state vectors are treated as points in a special manifold with a fiber bundle structure. The set of state vectors is thus realized as vector fields defined as the cross-sections of the fiber bundle. The transition of the system from one state to another is thus translated as the connectivity of points in a topological space through a mechanism of parallel transport. This movement in a curved manifold is guided by a quantity called a connection. Mathematically, the connection is defined as a one-form, which is a differential form of rank one, that is a vector field of the dual space of the tangent bundle of the manifold.

A topological index can be defined by integrating a differential form of appropriate rank over the entire manifold of parameters to characterize the geometric features of the fiber bundle. The two models considered exhibit different levels of sophistication in characterizing the nature governing the set of physical states. In the general two-level model, the topological properties are evident in the singularity of the state vector fields defined over the entire manifold of the model parameters. In contrast, for the SSH model, greater sophistication is required to discern the presence of a topological structure in a set of state vectors at an intermediate level of analysis. Additionally, we show that the set of state vectors does not always exhibit the fiber bundle structure directly over the Brillouin zone, contrary to intuitive thinking. Therefore, semiconducting atomic chains can be classified by the topological properties of two fiber bundles with the same total and fiber spaces but different base spaces.

This brief article cannot encompass all aspects of topology and its applications in physics, nor can it address the diverse range of physical problems to which topology can be applied. However, we hope that it has provided insight into the use of topological theories to characterize the fundamental structure of material phases in the quantum realm.

\bibliographystyle{apsrev4-1}
\bibliography{bibliography}

\begin{thebibliography}{26}%
\makeatletter
\providecommand \@ifxundefined [1]{%
 \@ifx{#1\undefined}
}%
\providecommand \@ifnum [1]{%
 \ifnum #1\expandafter \@firstoftwo
 \else \expandafter \@secondoftwo
 \fi
}%
\providecommand \@ifx [1]{%
 \ifx #1\expandafter \@firstoftwo
 \else \expandafter \@secondoftwo
 \fi
}%
\providecommand \natexlab [1]{#1}%
\providecommand \enquote  [1]{``#1''}%
\providecommand \bibnamefont  [1]{#1}%
\providecommand \bibfnamefont [1]{#1}%
\providecommand \citenamefont [1]{#1}%
\providecommand \href@noop [0]{\@secondoftwo}%
\providecommand \href [0]{\begingroup \@sanitize@url \@href}%
\providecommand \@href[1]{\@@startlink{#1}\@@href}%
\providecommand \@@href[1]{\endgroup#1\@@endlink}%
\providecommand \@sanitize@url [0]{\catcode `\\12\catcode `\$12\catcode
  `\&12\catcode `\#12\catcode `\^12\catcode `\_12\catcode `\%12\relax}%
\providecommand \@@startlink[1]{}%
\providecommand \@@endlink[0]{}%
\providecommand \url  [0]{\begingroup\@sanitize@url \@url }%
\providecommand \@url [1]{\endgroup\@href {#1}{\urlprefix }}%
\providecommand \urlprefix  [0]{URL }%
\providecommand \Eprint [0]{\href }%
\providecommand \doibase [0]{http://dx.doi.org/}%
\providecommand \selectlanguage [0]{\@gobble}%
\providecommand \bibinfo  [0]{\@secondoftwo}%
\providecommand \bibfield  [0]{\@secondoftwo}%
\providecommand \translation [1]{[#1]}%
\providecommand \BibitemOpen [0]{}%
\providecommand \bibitemStop [0]{}%
\providecommand \bibitemNoStop [0]{.\EOS\space}%
\providecommand \EOS [0]{\spacefactor3000\relax}%
\providecommand \BibitemShut  [1]{\csname bibitem#1\endcsname}%
\let\auto@bib@innerbib\@empty
\bibitem [{\citenamefont {Cayssol}\ and\ \citenamefont
  {Fuchs}(2021)}]{Cayssol_2021}%
  \BibitemOpen
  \bibfield  {author} {\bibinfo {author} {\bibfnamefont {J.}~\bibnamefont
  {Cayssol}}\ and\ \bibinfo {author} {\bibfnamefont {J.~N.}\ \bibnamefont
  {Fuchs}},\ }\href {\doibase https://doi.org/10.1088/2515-7639/abf0b5}
  {\bibfield  {journal} {\bibinfo  {journal} {J. Phys.: Materials}\ }\textbf
  {\bibinfo {volume} {4}},\ \bibinfo {pages} {034007} (\bibinfo {year}
  {2021})}\BibitemShut {NoStop}%
\bibitem [{\citenamefont {Price}\ \emph {et~al.}(2022)\citenamefont {Price},
  \citenamefont {Chong}, \citenamefont {Khanikaev}, \citenamefont {Schomerus},
  \citenamefont {Maczewsky}, \citenamefont {Kremer}, \citenamefont {Heinrich},
  \citenamefont {Szameit}, \citenamefont {Zilberberg}, \citenamefont {Yang},
  \citenamefont {Zhang}, \citenamefont {Alu}, \citenamefont {Thomale},
  \citenamefont {Carusotto}, \citenamefont {St-Jean}, \citenamefont {Amo},
  \citenamefont {Dutt}, \citenamefont {Yuan}, \citenamefont {Fan},
  \citenamefont {Yin}, \citenamefont {Peng}, \citenamefont {Ozawa},\ and\
  \citenamefont {Blanco-Redondo}}]{Price_2022}%
  \BibitemOpen
  \bibfield  {author} {\bibinfo {author} {\bibfnamefont {H.}~\bibnamefont
  {Price}}, \bibinfo {author} {\bibfnamefont {Y.}~\bibnamefont {Chong}},
  \bibinfo {author} {\bibfnamefont {A.}~\bibnamefont {Khanikaev}}, \bibinfo
  {author} {\bibfnamefont {H.}~\bibnamefont {Schomerus}}, \bibinfo {author}
  {\bibfnamefont {L.~J.}\ \bibnamefont {Maczewsky}}, \bibinfo {author}
  {\bibfnamefont {M.}~\bibnamefont {Kremer}}, \bibinfo {author} {\bibfnamefont
  {M.}~\bibnamefont {Heinrich}}, \bibinfo {author} {\bibfnamefont
  {A.}~\bibnamefont {Szameit}}, \bibinfo {author} {\bibfnamefont
  {O.}~\bibnamefont {Zilberberg}}, \bibinfo {author} {\bibfnamefont
  {Y.}~\bibnamefont {Yang}}, \bibinfo {author} {\bibfnamefont {B.}~\bibnamefont
  {Zhang}}, \bibinfo {author} {\bibfnamefont {A.}~\bibnamefont {Alu}}, \bibinfo
  {author} {\bibfnamefont {R.}~\bibnamefont {Thomale}}, \bibinfo {author}
  {\bibfnamefont {I.}~\bibnamefont {Carusotto}}, \bibinfo {author}
  {\bibfnamefont {P.}~\bibnamefont {St-Jean}}, \bibinfo {author} {\bibfnamefont
  {A.}~\bibnamefont {Amo}}, \bibinfo {author} {\bibfnamefont {A.}~\bibnamefont
  {Dutt}}, \bibinfo {author} {\bibfnamefont {L.}~\bibnamefont {Yuan}}, \bibinfo
  {author} {\bibfnamefont {S.}~\bibnamefont {Fan}}, \bibinfo {author}
  {\bibfnamefont {X.}~\bibnamefont {Yin}}, \bibinfo {author} {\bibfnamefont
  {C.}~\bibnamefont {Peng}}, \bibinfo {author} {\bibfnamefont {T.}~\bibnamefont
  {Ozawa}}, \ and\ \bibinfo {author} {\bibfnamefont {A.}~\bibnamefont
  {Blanco-Redondo}},\ }\href {\doibase
  https://doi.org/10.1088/2515-7647/ac4ee4} {\bibfield  {journal} {\bibinfo
  {journal} {J. Phys. Photonics}\ }\textbf {\bibinfo {volume} {4}},\ \bibinfo
  {pages} {032501} (\bibinfo {year} {2022})}\BibitemShut {NoStop}%
\bibitem [{\citenamefont {Lan}\ \emph {et~al.}(2022)\citenamefont {Lan},
  \citenamefont {Chen}, \citenamefont {Gao}, \citenamefont {Zhang},\ and\
  \citenamefont {Sha}}]{Lan_2022}%
  \BibitemOpen
  \bibfield  {author} {\bibinfo {author} {\bibfnamefont {Z.}~\bibnamefont
  {Lan}}, \bibinfo {author} {\bibfnamefont {M.~L.~N.}\ \bibnamefont {Chen}},
  \bibinfo {author} {\bibfnamefont {F.}~\bibnamefont {Gao}}, \bibinfo {author}
  {\bibfnamefont {S.}~\bibnamefont {Zhang}}, \ and\ \bibinfo {author}
  {\bibfnamefont {W.~E.~I.}\ \bibnamefont {Sha}},\ }\href {\doibase
  https://doi.org/10.1016/j.revip.2022.100076} {\bibfield  {journal} {\bibinfo
  {journal} {Rev. Phys.}\ }\textbf {\bibinfo {volume} {9}},\ \bibinfo {pages}
  {100076} (\bibinfo {year} {2022})}\BibitemShut {NoStop}%
\bibitem [{\citenamefont {Bernevig}\ and\ \citenamefont
  {Hughes}(2013)}]{Bernevig_2013}%
  \BibitemOpen
  \bibfield  {author} {\bibinfo {author} {\bibfnamefont {B.~A.}\ \bibnamefont
  {Bernevig}}\ and\ \bibinfo {author} {\bibfnamefont {T.~L.}\ \bibnamefont
  {Hughes}},\ }\href@noop {} {\emph {\bibinfo {title} {Topological Insulators
  and Topological superconductors}}}\ (\bibinfo  {publisher} {Princeton
  University Press},\ \bibinfo {year} {2013})\BibitemShut {NoStop}%
\bibitem [{\citenamefont {von Klitzing}\ \emph {et~al.}(1980)\citenamefont {von
  Klitzing}, \citenamefont {Dorda},\ and\ \citenamefont
  {Pepper}}]{Klitzing_1980}%
  \BibitemOpen
  \bibfield  {author} {\bibinfo {author} {\bibfnamefont {K.}~\bibnamefont {von
  Klitzing}}, \bibinfo {author} {\bibfnamefont {G.}~\bibnamefont {Dorda}}, \
  and\ \bibinfo {author} {\bibfnamefont {M.}~\bibnamefont {Pepper}},\ }\href
  {\doibase https://doi.org/10.1103/PhysRevLett.45.494} {\bibfield  {journal}
  {\bibinfo  {journal} {Phys. Rev. Lett.}\ }\textbf {\bibinfo {volume} {45}},\
  \bibinfo {pages} {494} (\bibinfo {year} {1980})}\BibitemShut {NoStop}%
\bibitem [{\citenamefont {Thouless}\ \emph {et~al.}(1982)\citenamefont
  {Thouless}, \citenamefont {Kohmoto}, \citenamefont {Nightingal},\ and\
  \citenamefont {den Nijs}}]{Thouless_1982}%
  \BibitemOpen
  \bibfield  {author} {\bibinfo {author} {\bibfnamefont {D.~J.}\ \bibnamefont
  {Thouless}}, \bibinfo {author} {\bibfnamefont {M.}~\bibnamefont {Kohmoto}},
  \bibinfo {author} {\bibfnamefont {P.}~\bibnamefont {Nightingal}}, \ and\
  \bibinfo {author} {\bibfnamefont {M.}~\bibnamefont {den Nijs}},\ }\href
  {\doibase https://doi.org/10.1103/PhysRevLett.49.405} {\bibfield  {journal}
  {\bibinfo  {journal} {Phys. Rev. Lett.}\ }\textbf {\bibinfo {volume} {49}},\
  \bibinfo {pages} {405} (\bibinfo {year} {1982})}\BibitemShut {NoStop}%
\bibitem [{\citenamefont {A.Yu.Kitaev}(2003)}]{Kitaev_2003}%
  \BibitemOpen
  \bibfield  {author} {\bibinfo {author} {\bibnamefont {A.Yu.Kitaev}},\ }\href
  {\doibase https://doi.org/10.1016/S0003-4916(02)00018-0} {\bibfield
  {journal} {\bibinfo  {journal} {Anal. Phys.}\ }\textbf {\bibinfo {volume}
  {303}},\ \bibinfo {pages} {2} (\bibinfo {year} {2003})}\BibitemShut {NoStop}%
\bibitem [{\citenamefont {Gilbert}(2021)}]{Gilbert_2021}%
  \BibitemOpen
  \bibfield  {author} {\bibinfo {author} {\bibfnamefont {M.~J.}\ \bibnamefont
  {Gilbert}},\ }\href {\doibase https://doi.org/10.1038/s42005-021-00569-5}
  {\bibfield  {journal} {\bibinfo  {journal} {Commun. Phys. (Nature)}\ }\textbf
  {\bibinfo {volume} {4}},\ \bibinfo {pages} {70} (\bibinfo {year}
  {2021})}\BibitemShut {NoStop}%
\bibitem [{wol()}]{wolfram_euler}%
  \BibitemOpen
  \href@noop {} {}\bibinfo {howpublished}
  {\url{https://mathworld.wolfram.com/KoenigsbergBridgeProblem.html}}\BibitemShut
  {NoStop}%
\bibitem [{bri({\natexlab{a}})}]{britannica_poincare}%
  \BibitemOpen
  \href@noop {} {}\bibinfo {howpublished}
  {\url{https://www.britannica.com/science/topology/History-of-topology}}
  ({\natexlab{a}})\BibitemShut {NoStop}%
\bibitem [{bri({\natexlab{b}})}]{britanica_gauss_law}%
  \BibitemOpen
  \href@noop {} {}\bibinfo {howpublished}
  {\url{https://www.britannica.com/science/Gausss-law}}
  ({\natexlab{b}})\BibitemShut {NoStop}%
\bibitem [{bri({\natexlab{c}})}]{britanica_ampere_law}%
  \BibitemOpen
  \href@noop {} {}\bibinfo {howpublished}
  {\url{https://www.britannica.com/science/Amperes-law}}
  ({\natexlab{c}})\BibitemShut {NoStop}%
\bibitem [{\citenamefont {Dirac}(1931)}]{Dirac_1931}%
  \BibitemOpen
  \bibfield  {author} {\bibinfo {author} {\bibfnamefont {P.~A.~M.}\
  \bibnamefont {Dirac}},\ }\href {\doibase
  https://doi.org/10.1098/rspa.1931.0130} {\bibfield  {journal} {\bibinfo
  {journal} {Proc. Roy. Soc. London}\ }\textbf {\bibinfo {volume} {A133}},\
  \bibinfo {pages} {60} (\bibinfo {year} {1931})}\BibitemShut {NoStop}%
\bibitem [{\citenamefont {Wu}\ and\ \citenamefont
  {Yang}(1975{\natexlab{a}})}]{Wu_1975a}%
  \BibitemOpen
  \bibfield  {author} {\bibinfo {author} {\bibfnamefont {T.~T.}\ \bibnamefont
  {Wu}}\ and\ \bibinfo {author} {\bibfnamefont {C.~N.}\ \bibnamefont {Yang}},\
  }\href {\doibase https://doi.org/10.1103/PhysRevD.12.3843} {\bibfield
  {journal} {\bibinfo  {journal} {Phys. Rev. D}\ }\textbf {\bibinfo {volume}
  {12}},\ \bibinfo {pages} {3843} (\bibinfo {year}
  {1975}{\natexlab{a}})}\BibitemShut {NoStop}%
\bibitem [{\citenamefont {Wu}\ and\ \citenamefont
  {Yang}(1975{\natexlab{b}})}]{Wu_1975b}%
  \BibitemOpen
  \bibfield  {author} {\bibinfo {author} {\bibfnamefont {T.~T.}\ \bibnamefont
  {Wu}}\ and\ \bibinfo {author} {\bibfnamefont {C.~N.}\ \bibnamefont {Yang}},\
  }\href {\doibase https://doi.org/10.1103/PhysRevD.12.3845} {\bibfield
  {journal} {\bibinfo  {journal} {Phys. Rev. D}\ }\textbf {\bibinfo {volume}
  {12}},\ \bibinfo {pages} {3845} (\bibinfo {year}
  {1975}{\natexlab{b}})}\BibitemShut {NoStop}%
\bibitem [{\citenamefont {Aharonov}\ and\ \citenamefont
  {Bohm}(1959)}]{Aharonov_1959}%
  \BibitemOpen
  \bibfield  {author} {\bibinfo {author} {\bibfnamefont {Y.}~\bibnamefont
  {Aharonov}}\ and\ \bibinfo {author} {\bibfnamefont {D.}~\bibnamefont
  {Bohm}},\ }\href {\doibase https://doi.org/10.1103/PhysRev.115.485}
  {\bibfield  {journal} {\bibinfo  {journal} {Phys. Rev.}\ }\textbf {\bibinfo
  {volume} {115}},\ \bibinfo {pages} {485} (\bibinfo {year}
  {1959})}\BibitemShut {NoStop}%
\bibitem [{\citenamefont {Chambers}(1960)}]{Chambers_1960}%
  \BibitemOpen
  \bibfield  {author} {\bibinfo {author} {\bibfnamefont {R.~G.}\ \bibnamefont
  {Chambers}},\ }\href {\doibase https://doi.org/10.1103/PhysRevLett.5.3}
  {\bibfield  {journal} {\bibinfo  {journal} {Phys. Rev. Lett.}\ }\textbf
  {\bibinfo {volume} {5}},\ \bibinfo {pages} {3} (\bibinfo {year}
  {1960})}\BibitemShut {NoStop}%
\bibitem [{\citenamefont {Berry}()}]{Berry_1984}%
  \BibitemOpen
  \bibfield  {author} {\bibinfo {author} {\bibfnamefont {M.~V.}\ \bibnamefont
  {Berry}},\ }\href {\doibase https://doi.org/10.1098/rspa.1984.0023}
  {\bibfield  {journal} {\bibinfo  {journal} {Proc. Roy. Soc. A}\ }\textbf
  {\bibinfo {volume} {392}},\ \bibinfo {pages} {45}}\BibitemShut {NoStop}%
\bibitem [{\citenamefont {Berry}(1990)}]{Berry_1990}%
  \BibitemOpen
  \bibfield  {author} {\bibinfo {author} {\bibfnamefont {M.~V.}\ \bibnamefont
  {Berry}},\ }\href@noop {} {\bibfield  {journal} {\bibinfo  {journal} {Phys.
  Today}\ ,\ \bibinfo {pages} {34}} (\bibinfo {year} {1990})}\BibitemShut
  {NoStop}%
\bibitem [{\citenamefont {Resta}(2000)}]{Resta_2000}%
  \BibitemOpen
  \bibfield  {author} {\bibinfo {author} {\bibnamefont {Resta}},\ }\href
  {\doibase https://doi.org/10.1088/0953-8984/12/9/201} {\bibfield  {journal}
  {\bibinfo  {journal} {J. Phys. Cond. Matt.}\ }\textbf {\bibinfo {volume}
  {12}},\ \bibinfo {pages} {R107} (\bibinfo {year} {2000})}\BibitemShut
  {NoStop}%
\bibitem [{\citenamefont {Shapere}\ and\ \citenamefont
  {(ed.)}(1989)}]{Shapere_1989}%
  \BibitemOpen
  \bibfield  {author} {\bibinfo {author} {\bibfnamefont {A.}~\bibnamefont
  {Shapere}}\ and\ \bibinfo {author} {\bibfnamefont {F.~W.}\ \bibnamefont
  {(ed.)}},\ }\href@noop {} {\emph {\bibinfo {title} {Geometric phase in
  physics}}}\ (\bibinfo  {publisher} {Singapore: World Scientific},\ \bibinfo
  {year} {1989})\BibitemShut {NoStop}%
\bibitem [{\citenamefont {Asb\'oth}\ \emph {et~al.}(2016)\citenamefont
  {Asb\'oth}, \citenamefont {Oroszl\'any},\ and\ \citenamefont
  {P\'alyi}}]{Asboth_2016}%
  \BibitemOpen
  \bibfield  {author} {\bibinfo {author} {\bibfnamefont {J.~K.}\ \bibnamefont
  {Asb\'oth}}, \bibinfo {author} {\bibfnamefont {L.}~\bibnamefont
  {Oroszl\'any}}, \ and\ \bibinfo {author} {\bibfnamefont {A.}~\bibnamefont
  {P\'alyi}},\ }\href {\doibase 10.1007/978-3-319-25607-8} {\emph {\bibinfo
  {title} {A short course on topological insulators: Band structure and Edge
  states in one and two dimensions}}}\ (\bibinfo  {publisher} {Springer},\
  \bibinfo {year} {2016})\BibitemShut {NoStop}%
\bibitem [{\citenamefont {Nakahara}(2003)}]{Nakahara_2003}%
  \BibitemOpen
  \bibfield  {author} {\bibinfo {author} {\bibfnamefont {M.}~\bibnamefont
  {Nakahara}},\ }\href@noop {} {\emph {\bibinfo {title} {Geometry, Topology and
  Physics}}},\ \bibinfo {edition} {2nd}\ ed.\ (\bibinfo  {publisher} {IOP
  Publishing Ltd.},\ \bibinfo {year} {2003})\BibitemShut {NoStop}%
\bibitem [{\citenamefont {Hassani}(1999)}]{Hassani_1999}%
  \BibitemOpen
  \bibfield  {author} {\bibinfo {author} {\bibfnamefont {S.}~\bibnamefont
  {Hassani}},\ }\href@noop {} {\emph {\bibinfo {title} {Mathematical Physics: A
  modern introduction ot its foundations}}}\ (\bibinfo  {publisher}
  {Springer},\ \bibinfo {year} {1999})\BibitemShut {NoStop}%
\bibitem [{\citenamefont {Simon}(2021)}]{Simon_2021}%
  \BibitemOpen
  \bibfield  {author} {\bibinfo {author} {\bibfnamefont {D.~S.}\ \bibnamefont
  {Simon}},\ }\href {\doibase 10.1088/978-0-7503-3471-6} {\emph {\bibinfo
  {title} {Topology in Optics}}}\ (\bibinfo  {publisher} {IOP Publishing
  Ltm.},\ \bibinfo {year} {2021})\BibitemShut {NoStop}%
\bibitem [{\citenamefont {El-Batanouny}(2020)}]{Micheal_2020}%
  \BibitemOpen
  \bibfield  {author} {\bibinfo {author} {\bibfnamefont {M.}~\bibnamefont
  {El-Batanouny}},\ }\href {\doibase 10.1017/9781108691291} {\emph {\bibinfo
  {title} {Advanced quantum condensed matter physics: one-body, many-body, and
  topological perspectives}}}\ (\bibinfo  {publisher} {Cambridge Univ.
  Press.},\ \bibinfo {year} {2020})\BibitemShut {NoStop}%
\end{thebibliography}%

\end{document}